\documentclass[fleqn,12pt,twoside]{article}
\usepackage[headings]{espcrc1}

\readRCS
$Id: semistructured.tex,v 1.5 2008/08/08 10:34:56 bradler Exp $
\usepackage{graphicx}
\usepackage[figuresright]{rotating}

\usepackage{cite}
\usepackage{amsmath}
\usepackage{amsthm}
\usepackage{subfigure}

\newtheorem{theorem}{Theorem}

\title{\vspace{-2,5cm}\normalsize Technical Report TUD-CS-2008-103,  \hfill ~ \\
Technische Universit\"at Darmstadt, 08.08.2008 \hfill ~ \\
\large \vspace{3,3cm}\textbf{Optimally Efficient Prefix Search and\\
 Multicast in Structured P2P Networks}\\~\\
\normalsize 
Technical Report No. TUD-CS-2008-103\\ \normalsize ~\\
Telecooperation Report No. TK-01/08, \\The Technical Reports Series of the
TK Research Division,\\ Technische Universit\"at Darmstadt
}

\author{D. Bradler\address[TK]{Telecooperation Group, 
        University of Technology Darmstadt, \\ 
        Darmstadt, Germany}%
        \thanks{this research is funded by DFG},
        J. Kangasharju\address{Department of Computer Science, University of Helsinki, Finnland},
        M. M\"uhlh\"auser\addressmark[TK]{}
}

\runtitle{Optimally Efficient Prefix Search and Multicast in
  Structured Peer-to-Peer Networks}
\runauthor{D. Bradler, J. Kangasharju, M. M\"uhlh\"auser}

\begin{document}

\maketitle

\section*{Abstract}

  Searching in P2P networks is fundamental to all overlay networks.
  P2P networks based on Distributed Hash Tables (DHT) are optimized for single key lookups, whereas
  unstructured networks offer more complex queries at the cost of
  increased traffic and uncertain success rates.  Our Distributed Tree
  Construction (DTC) approach enables structured P2P networks to
  perform prefix search, range queries, and multicast in an optimal
  way. It achieves this by creating a spanning tree over the peers in
  the search area, using only information available locally on each
  peer. Because DTC creates a spanning tree, it can query all the
  peers in the search area with a minimal number of messages.
  Furthermore, we show that the tree depth has the same upper bound as
  a regular DHT lookup which in turn guarantees fast and responsive
  runtime behavior. By placing objects with a region quadtree, we can
  perform a prefix search or a range query in a freely selectable area
  of the DHT. Our DTC algorithm is DHT-agnostic and works with most
  existing DHTs.  We evaluate the performance of DTC over several DHTs
  by comparing the performance to existing application-level
  multicast solutions, we show that DTC sends 30--250\% fewer messages
  than common solutions.

\section{Introduction}
\label{sec:introduction}
Peer-to-peer networks are typically divided into \emph{unstructured}
and \emph{structured} networks, depending on how the overlay is
constructed and how content is placed in the network. In an
unstructured network, nodes are free to choose their overlay neighbors
and are free to offer any content they want. In contrast, in a
structured network, the structuring mechanism (typically one or more
hash functions) uniquely determine the location of a peer in the
overlay and its neighbors, as well as the placement of content on
peers. 

Both these kinds of networks have their strengths and weaknesses.
Unstructured networks are based on \emph{searching} for content, which
allows us to use complex queries for determining which objects match
which requests. However, this comes at the expense of having to flood
the search query through the network which causes significant amount
of network traffic. The original Gnutella network was a completely
flat overlay which relied on locating objects by flooding the network
up to a given time-to-live (TTL). Newer unstructured networks, like Kazaa and
eDonkey, build a two-tier hierarchy with ordinary peers and
superpeers. Ordinary peers connect to one superpeer and the superpeers
build a Gnutella-like overlay between them. Although such a
hierarchical network scales better than the plain Gnutella network, we
still cannot guarantee that an object can be found and the amount of
network traffic caused by flooding between the superpeers can still be
high. A structured network, on the other hand, clearly defines the overlay
structure and object placement through hash functions. The main
advantage is that it allows for very efficient key-value lookups,
similar to traditional hash tables. However, because the content is
placed with hash functions, real search queries are not feasible in
DHTs. For example, it is not feasible to query a DHT for all objects
whose name begins with ``Foo''. This would usually require asking
\emph{every peer} whether it has any matching objects.

We tackle the problem of implementing a prefix search on a structured
P2P network. Our algorithm selects a subset of the overlay network,
and creates a spanning tree for that subset rooted at any of the peers
in that set. Typical applications which benefit from prefix search are
all systems which need to deal with structured data. Structured data
is very commonly used in many applications. For example storing data
based on geographical location (e.g., tourist information) or any kind
of classification systems lend themselves readily to our algorithm.

The more attributes are available in the user provided search term
(i.e., the more precise the user's query), the less nodes will be
queried. In the case of a user performing a very precise search, the
query will reduce itself to a standard DHT key lookup.  Because we
create a spanning tree, our solution is not limited to implementing
prefix search; other applications, such as multicast and broadcast,
can also be implemented with our algorithm. Multicast (and also
broadcast) is extremely useful in cases where the DHT is built
according to some specific criteria (as opposed to a standard hash
function; see Section~\ref{sec:hashing}), since it gives us the
possibility of reaching a given set of nodes with minimal overhead.

%% Locations 
%% are structured data, e.g. Asia$\xrightarrow{}$China$\xrightarrow{}$Bejing$\xrightarrow{}$Badaling Great Wall.
%% Movies and music can be stored as structured data as well, 
%% e.g. Country$\xrightarrow{}$USA$\xrightarrow{}$Carrie Underwood$\xrightarrow{}$So small. There are a lot of more  
%% application areas which utilize structured data sets. 

The key feature of our algorithm is that it creates a spanning tree
\emph{without any communication between the peers}, using only local
information available to every participating peer. Because it
is a tree, we are guaranteed that only the minimum number of messages
needs to be sent. These key features set our algorithm apart from
previous work, such as the application-level multicast proposed by
Ratnasamy et al.~\cite{ratnasamy2001alm} or
SplitStream~\cite{CastroM:SplitStream}.  Previous work typically
either has a high number of duplicate messages, requires additional
coordination traffic or needs a second overlay network. Additionally range and 
origin of the spanning tree can freely be set to any value and any peer.

Our distributed tree construction algorithm (DTC) presented in
Section~\ref{sec:flooding_algorithm} has none of the shortcomings of
the previous algorithms. It can be used on top of most existing DHTs
and it works using information available at each peer about the peer's
neighbors. As we show in Section~\ref{sec:flooding_algorithm}, this
standard information maintained by the DHT is sufficient for spanning
a tree.

As our evaluation shows, our DTC algorithm yields optimal performance
in terms of messages sent. We also show that the overhead of duplicate
messages in existing solutions is at least 30\% but can in several
cases be up to 250\%.

This paper is organized as follows. In
Section~\ref{sec:flooding_algorithm}, we present how we construct the
distributed spanning tree and prove its properties.
Section~\ref{sec:hashing} shows an example of how to build a prefix
search using hash functions and our tree construction algorithm.
Section~\ref{sec:evaluation} evaluates our algorithm on different DHTs
and compares its performance against existing algorithms. In
Section~\ref{sec:robustness-security}, we discuss the robustness of
our algorithm and present mechanisms for improving its resilience
against malicious peers. Section~\ref{sec:related-work} discusses
related work. Finally, Section~\ref{sec:conclusion} concludes the
paper.

\section{Distributed Tree Construction}
\label{sec:flooding_algorithm}

In unstructured networks, more than 70\% of the messages are
redundant, even with a moderate TTL~\cite{jiang2003lef}.  This high
overhead is one of the main reasons for the poor scalability of
unstructured networks. With our \emph{distributed tree construction}
algorithm (DTC), we are able to eliminate this overhead. Note that DTC
builds on top of a standard, structured overlay network. We first
discuss the requirements on the structured overlay, and then
present the DTC algorithm with optimality proofs.  Finally, in
Section~\ref{sec:sample-applications} we discuss different
applications and how to build them using DTC.

The idea behind DTC is to build a spanning tree to connect all the
nodes we want to search. When a query is sent from the root, every
node in the tree receives it exactly once. The challenge lies in
constructing the tree without any overhead and using only local information
available in each node.

\subsection{Structured Overlay}
\label{sec:structured-overlay}

Our DTC algorithm works on any structured overlay (a.k.a. distributed
hash table) which fulfills the property that every node knows
\emph{all} of its immediate neighbors in the overlay hash space.
Networks like Chord~\cite{StoicaI:Chord}, CAN~\cite{RatnasamyS:CAN} and VoroNet~\cite{beaumont2007vso}
obviously fulfill this property. In case of Chord, the critical
information is knowing the successor and for CAN, knowing all
neighbors in all coordinate directions. In case of
Pastry~\cite{RowstronA:Pastry} the condition is fulfilled since the
leaf sets of all nodes always contain the closest neighbors in the
hash space.  Although Tapestry~\cite{ZhaoB:TapestryJ} is very similar
to Pastry, it does not have the equivalent of the leaf set and thus
might not be suitable without modifications.
Kademlia~\cite{MaymounkovP:Kademlia} also fulfills the required
condition, since the buckets for the shorter distances contain the
closest nodes in the hash space and they should be complete.

In the remainder of this paper, we consider only Chord and CAN as
overlay networks. Although the principle of DTC in both networks is
the same, differences in the overlay structures lead to
performance differences (see Section~\ref{sec:evaluation}).

\subsection{DTC Algorithm}
\label{sec:dtc-algorithm}

In the distributed tree construction algorithm, we start spanning the
tree from a point in the overlay and expand from that point through
the overlay according to the overlay routing. The area of the overlay
which the spanning tree is supposed to cover is explicitly defined.
The information about the root of the tree and the area are sufficient
to construct the spanning tree in a purely distributed manner. We will
now show how this can be done in Chord and CAN.

Note that even though the algorithm constructs a spanning tree, no
peer has a complete view of the tree. The tree is always constructed
on-demand by having the root send a message which constructs the tree
as it gets passed through the peers. Thus, when we say below that a
peer adds some other peers to the tree as its children, it means in
practice that the peer in question forwards a message to the other
peers.

\noindent
\textbf{Example: Chord}

In case of Chord, the area is an arc on the Chord ring and the root of
the tree is the first node on the arc. The simplest solution is simply
to add nodes to the tree along the chain of successors until the end
of the arc has been reached. However, this is not very efficient for
large areas. Instead, we should use the fingers to create shortcuts
and broaden the spanning tree. The root of the tree selects all of its
fingers that are in the area as its children. Each of them will
recursively perform the same operation until all peers in the area
(the arc) have been included in the tree. Any of the peers can easily
determine which of its fingers it should include, since it knows the
root and the length of the arc. The successor of the last point of the
arc\footnote{Often the first node after the arc.} \emph{is} part of
the tree, since the last points in the arc might contain objects which
are stored on their successor.

\noindent
\textbf{Example: CAN}

In case of CAN, the area is a convex area of the $d$-dimensional
coordinate space, with the root somewhere in this area. The
restriction to convex search areas is imposed by our algorithm.
Non-convex areas can be searched by splitting the search into
non-overlapping convex searches which cover the desired area. The root
first adds its immediate neighbors ($2d$ neighbors in a
$d$-dimensional CAN), which then continue adding their neighbors,
according to the rules defined below.  As in the Chord-case above, the
information about the root of the tree and the area it is supposed to
cover are available to the peers.  The area can be defined either with
simply the radius of the area, or by specifying for each dimension
separately how far the area reaches in that dimension.  As mentioned,
the only restriction on the search area is that it must be convex.

We assume every node knows the following:
\begin{itemize}
\item Size of the zone of each neighbor (maintained by standard CAN
  routines)
\item Root of the tree and the area it is supposed to cover (available
  in the message which is used to create the tree)
\end{itemize}

\begin{figure}[!tb]
  \centering
  \includegraphics[width=0.40\textwidth]{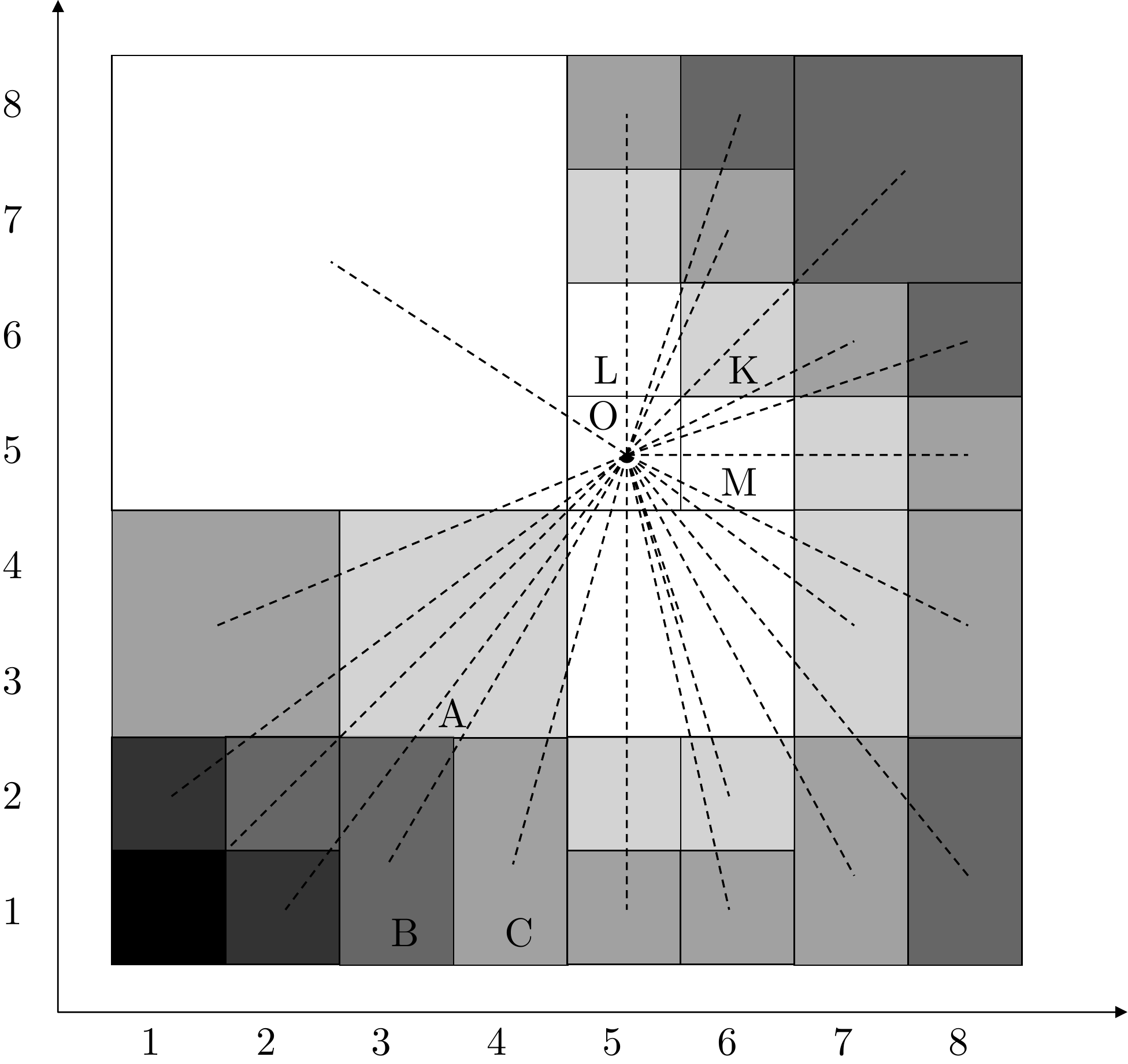}
  \caption{Example spanning tree of a CAN}
  \label{fig:flooding}
  \vskip -5mm
\end{figure}

Figure~\ref{fig:flooding} shows in an example a spanning tree of a two dimensional CAN. 
The tree is rooted at the white zone marked $O$ at coordinates (5,~5). The other white zones are
the children of the root, and the levels of the tree are shown in
increasingly darker shades of gray. The result is a spanning tree
consisting of all nodes within the CAN overlay.

The tree is constructed as follows.  When a peer $X$ receives the
query, it computes for every one of its neighbors the vector from the
center of the root's zone to the center of the neighbor's zone. If
that vector intersects the common border surface between $X$ and the
neighbor, then $X$ should add that neighbor as its child. The vector
has to intersect the common border surface between the two nodes; it
is not sufficient for the vector to pass through $X$'s zone. Consider
the third zone from the left on the bottom row in
Figure~\ref{fig:flooding}, marked $B$. This zone has one light gray
neighbor $A$ on the top and one medium gray $C$ neighbor to the right.
The vector from the root passes through both of these neighbors, but
the one on the right (marked $C$) is the parent of node $B$.

It is important to note that every node is able to compute the vectors
and determine whether it should add any of its neighbors as children
(and thus forward the message) by using only information available
locally through normal overlay communications.  No coordination
between nodes is needed, nor is any additional traffic generated.

\subsection{Proof of Optimality}
\label{sec:proof-optimality}

We prove the following properties of our DTC algorithm:
\begin{enumerate}

\item The DTC algorithm creates a spanning tree over the area
\item The depth of the tree is proportional to message complexity of the
  underlying DHT
\end{enumerate}

For the simple version of a Chord-based DTC (i.e., every node passes
the query to its successor), the first property is obvious. The proof
of the first property with fingers is also straight-forward and is
omitted for space reasons.

In the following, we will prove the properties for a CAN-based DTC. We
make the simplifying assumption that the overlay network is able to
heal itself under churn without loss of messages.  Note that this
assumption explicitly \emph{allows} churn as long as changes to the
overlay structure are performed in a locally atomic manner and no node
departs between receiving and forwarding a query. We also assume that
the area over which the tree is to be spanned is convex.

\begin{theorem}
  \label{thm:at-least-once}
  All nodes are added to the tree at least once.
\end{theorem}
\begin{proof}
  For every zone $Z$ in the convex area, there is a single vector
  which connects the center points of that zone with the zone of the
  root of the tree. Starting from zone $Z$, the vector determines $Y$,
  a neighbor of $Z$ who will add $Z$ as its child.  Considering zone
  $Y$, we can draw the vector between the center of $Y$ and the
  starting zone, which determines a zone $X$, neighbor of $Y$ which
  adds $Y$ as a child. Continuing in a similar manner, we arrive at
  the zone of the root of the tree. Thus, we are able to find a chain
  of zones which leads us from the root zone to zone $Z$. Thus,
  all nodes have at least one path from the root, i.e., are part of
  the tree.

  In some cases, it is possible that the vector between the
  root and a zone $Z$ does not pass through any direct CAN neighbor of
  $Z$. For example, in Figure~\ref{fig:flooding}, the vector between
  the root and the zone marked $K$ passes directly through the corner
  point of the two zones. Depending on how the ownership of edges is
  defined, it is possible that there is no neighbor through whose zone
  the vector passes on its way from root to $K$.  (Note that
  regardless of how the ownership of edges is defined, it is always
  possible to construct the zones such that this problem persists.) In
  general, this issue arises when two zones share up to $(d-2)$
  dimensions in a $d$-dimensional CAN (e.g., a point in 2-dimensional
  CAN and a point or a line in a 3-dimensional CAN). In this case, the
  forwarding algorithm does not reach all nodes. We have defined the
  following tie breaker for these cases.

  \textbf{The Tie Breaker:} We use the following rule for determining
  how to construct the tree in the above case. The two problematic
  zones differ in at least 2 and up to $d$ dimensions. We order the
  dimensions beforehand.  The forwarding path should be such that the
  smallest dimensions with differences are used first. The length of
  the tie breaker path will be the same as the number of dimensions in
  which the two problem zones differ (i.e., between 2 and $d$). Note
  that none of the nodes on the path would normally forward the query,
  but \emph{all of them} are able to compute locally that they are
  part of the tie breaker procedure and are able to perform their
  duties correctly. In the example of Figure~\ref{fig:flooding}, the
  tie breaker would mean that $M$ is the node responsible for adding
  $K$ as its child, since x-coordinate is considered before
  y-coordinate.
\end{proof}

\begin{theorem}
  \label{thm:at-most-once}
  All nodes are added to the tree at most once.
\end{theorem}
\begin{proof}
  We prove this by contradiction. If a node $A$ were to be added to
  the tree twice, this would imply that two of its neighbors would
  think that the vector between $A$ and the root passes through their
  zones. This is clearly impossible, since the responsibility is
  defined by the vector and the vector between $A$'s center point and
  the center of the root's zone intersects only one of the borders
  between $A$ and its neighbors. Thus, $A$ can be added to the tree by
  at most one of its neighbors.
\end{proof}

\begin{figure*}[!tb]
  \centering
  \mbox{
    \subfigure[Spanning tree overlaid on CAN]{\label{fig:flood-onlyarrows-64}\includegraphics[width=0.40\textwidth]{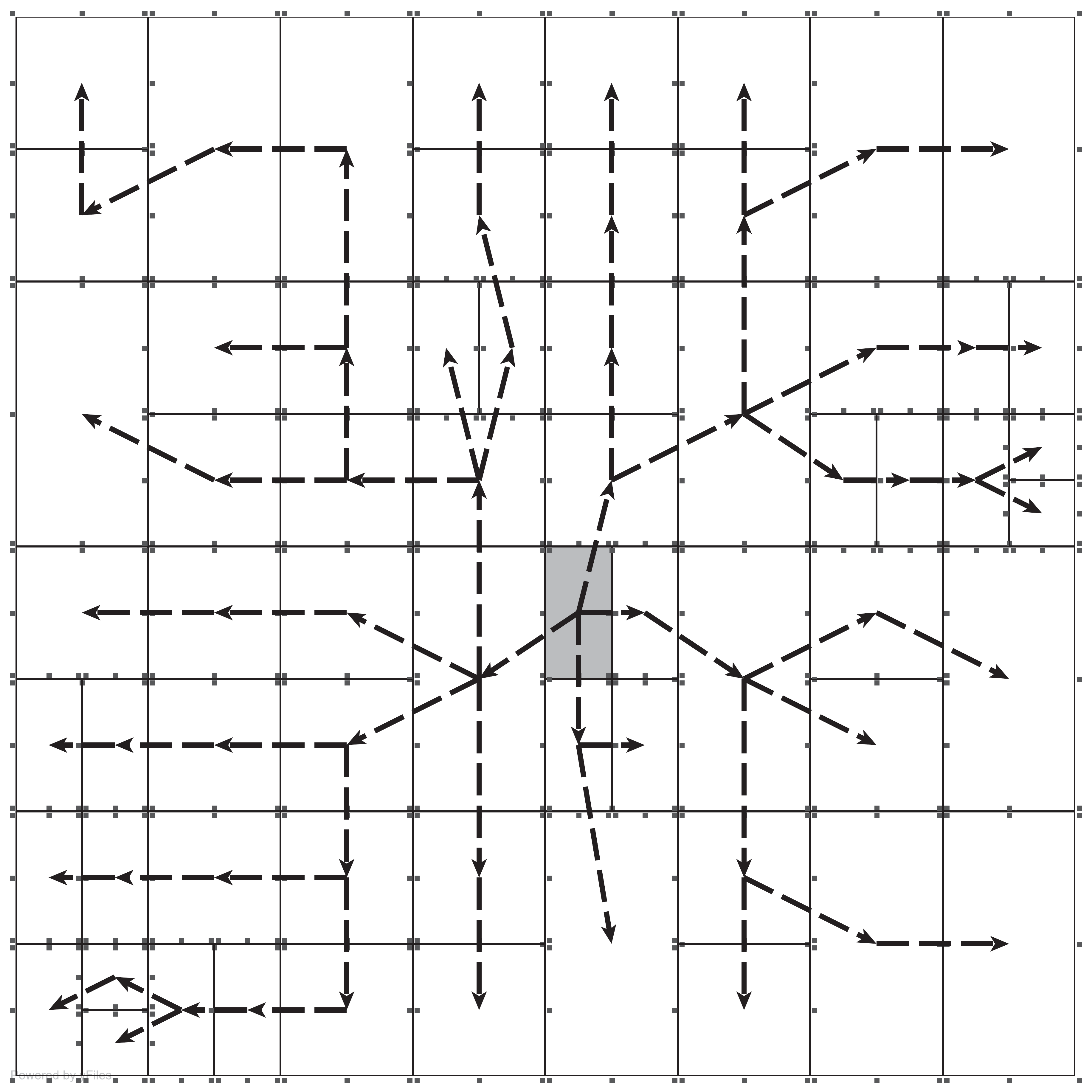}}
    \quad
    \subfigure[Spanning tree]{\label{fig:flood-ant-64}\includegraphics[width=0.40\textwidth]{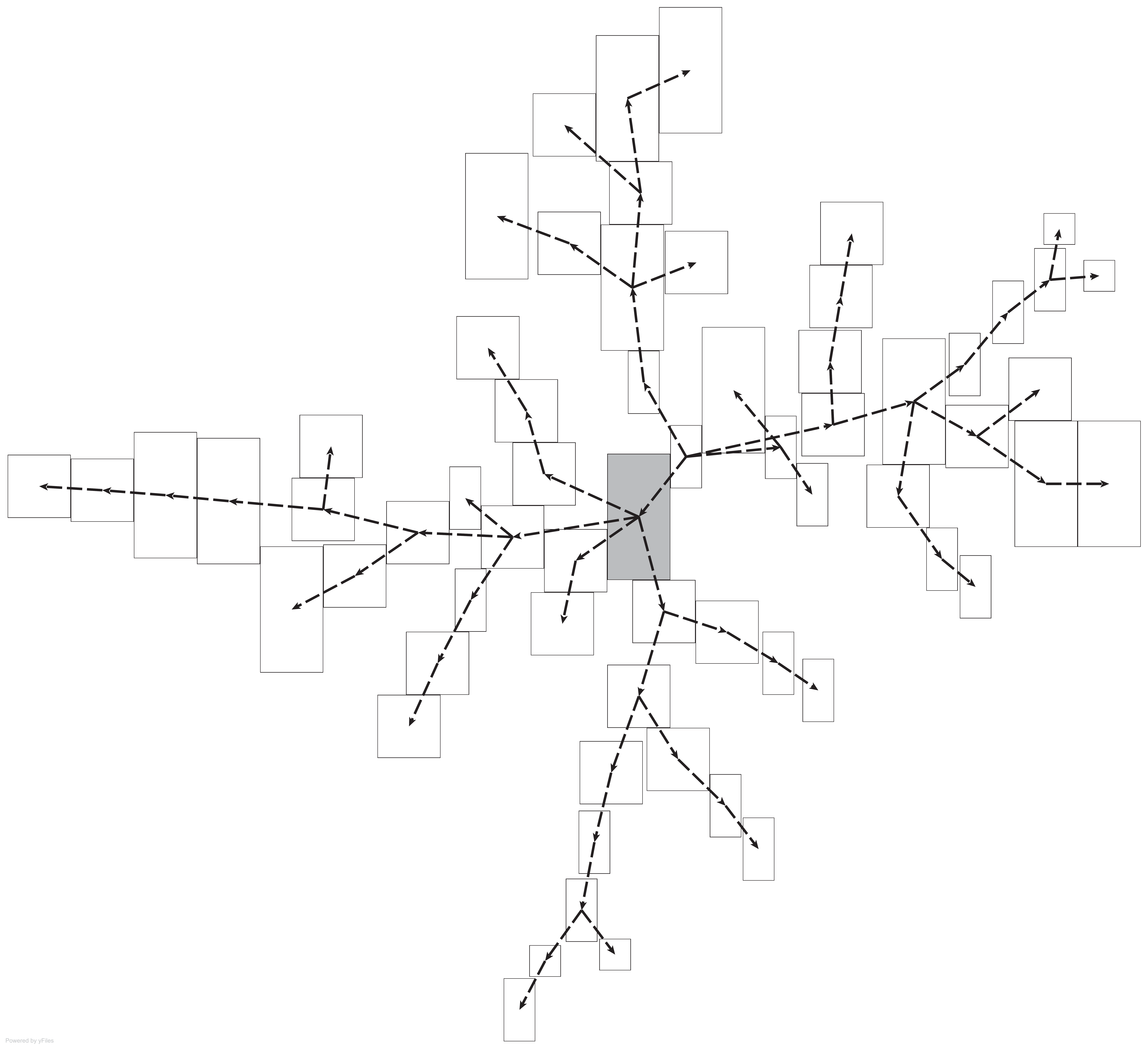}}
    }
  \caption{Example of DTC-constructed tree}
  \label{fig:example-dtc-constr}
  \vskip -5mm
\end{figure*}

Theorems~\ref{thm:at-least-once} and~\ref{thm:at-most-once} prove that
every node in the area is added to the tree exactly once, thus the DTC
algorithm creates a spanning tree rooted at the root zone and covering
all the nodes in the area. Note that the spanning tree is only
\emph{a} spanning tree; it might not be the minimal spanning tree (but
this property is not a requirement of the applications we are considering).

Figure~\ref{fig:example-dtc-constr} shows a larger example of how the
DTC algorithm constructs the spanning tree. In
Figure~\ref{fig:flood-onlyarrows-64}, we show the spanning tree
overlaid on the underlying CAN topology and show which peers add which
other peers as their children. Figure~\ref{fig:flood-ant-64} shows only the
resulting spanning tree.

\begin{theorem}
  Depth of the spanning tree is proportional to message complexity of
  the underlying DHT.
\end{theorem}

We prove this for both Chord- and CAN-based DTC.  In case of a
Chord-based DTC, we can construct the spanning tree using fingers, as
mentioned above. The links in the Chord-DTC spanning tree are
determined by the fingers and successor pointers of the nodes.

The claim of the theorem refers to the depth of the spanning tree.
Note that every path in the spanning tree is a legal DHT-routing path
between the root and the chosen node.

\begin{proof}[Proof for a Chord-based system]
  In a Chord-based system with fingers, the spanning tree is simply a
  mapping from the fingers and successor pointers to the nodes, and
  every path in the tree exactly corresponds to the routing path that the 
  standard Chord routing would take to reach that node. Hence, the
  depth of the tree is $O(\log(N))$.
\end{proof}

\begin{proof}[Proof for a CAN-based system]
  In a CAN-based system, every hop is also a legal CAN routing hop,
  however, not necessarily a hop that the standard greedy CAN routing
  would take in a given situation. The DTC algorithm always follows the
  vectors, but the greedy CAN routing might take shortcuts over large
  zones.  Nevertheless, the length of the path in the spanning tree is
  still $O(\sqrt[d]{n})$.
\end{proof}

\subsection{Sample Applications}
\label{sec:sample-applications}

The ability to construct a spanning tree from any point in the DHT is
very powerful, and allows us to develop many different kinds of
applications. We now discuss some of the applications which can be
built with DTC.

\noindent
\textbf{Prefix Search in DHTs}

As mentioned in the introduction, searching in DHTs is extremely
inefficient. With DTC and the hashing scheme from
Section~\ref{sec:hashing}, we are able to implement a prefix search
over a freely selectable prefix with only the minimum number of
messages needed. We achieve this through a slight modification of how
content is mapped on the nodes (see below) and by spanning a tree over
a pre-determined area.

As discussed in the introduction, many applications can benefit from
prefix searches. In particular, applications which use any kind of
structured data lend themselves readily to prefix searches. Structured
data is very common and easily maps to hierarchical concepts which are
used in many different applications.

\noindent
\textbf{Group Communication Primitives}

If the root sends a message along the tree (as is done during the tree
construction), then every node in the area will receive the message
exactly once. By tuning the area which the tree spans, we can easily
define different multicast groups and reach them with the minimum
number of messages. (As the comparison in Section~\ref{sec:evaluation}
shows, the application-level multicast on CAN~\cite{ratnasamy2001alm}
has significant overhead compared to our DTC-based approach.) It is
even possible to let the area be the complete hash space of the DHT,
in which case we have an optimal broadcast mechanism.

An important point to keep in mind when designing applications running
with DTC is whether there will be feedback to the root of the tree or
not. In other words, a search requires an answer, i.e., all the nodes
in the tree with matching content should answer. In contrast, a
multicast or a broadcast might not require any acknowledgement from
the receivers. The presence or absence of feedback is thus
application-dependent, and we will return to this issue in
Section~\ref{sec:robustness-security}. 

\section{Prefix Search in a DHT}
\label{sec:hashing}

We now present a way of mapping objects to peers in a DHT, such that
the DTC algorithm allows us to perform prefix searches on the DHT. We 
achieve this by placing all objects matching a
prefix in a certain area of the DHT and the prefix search corresponds
simply to spanning a tree with the DTC algorithm over this area. As an
example of how to implement a prefix search, we consider the case of a
$d$-dimensional CAN. (Note that other DHTs mentioned as candidates in
Section~\ref{sec:structured-overlay} are 1-dimensional, thus they are
simply special cases of the example.)

\subsection{Quad Trees}
\label{sec:quad-trees}

\begin{figure}[!tb]
  \centering
  \includegraphics[width=0.40\textwidth]{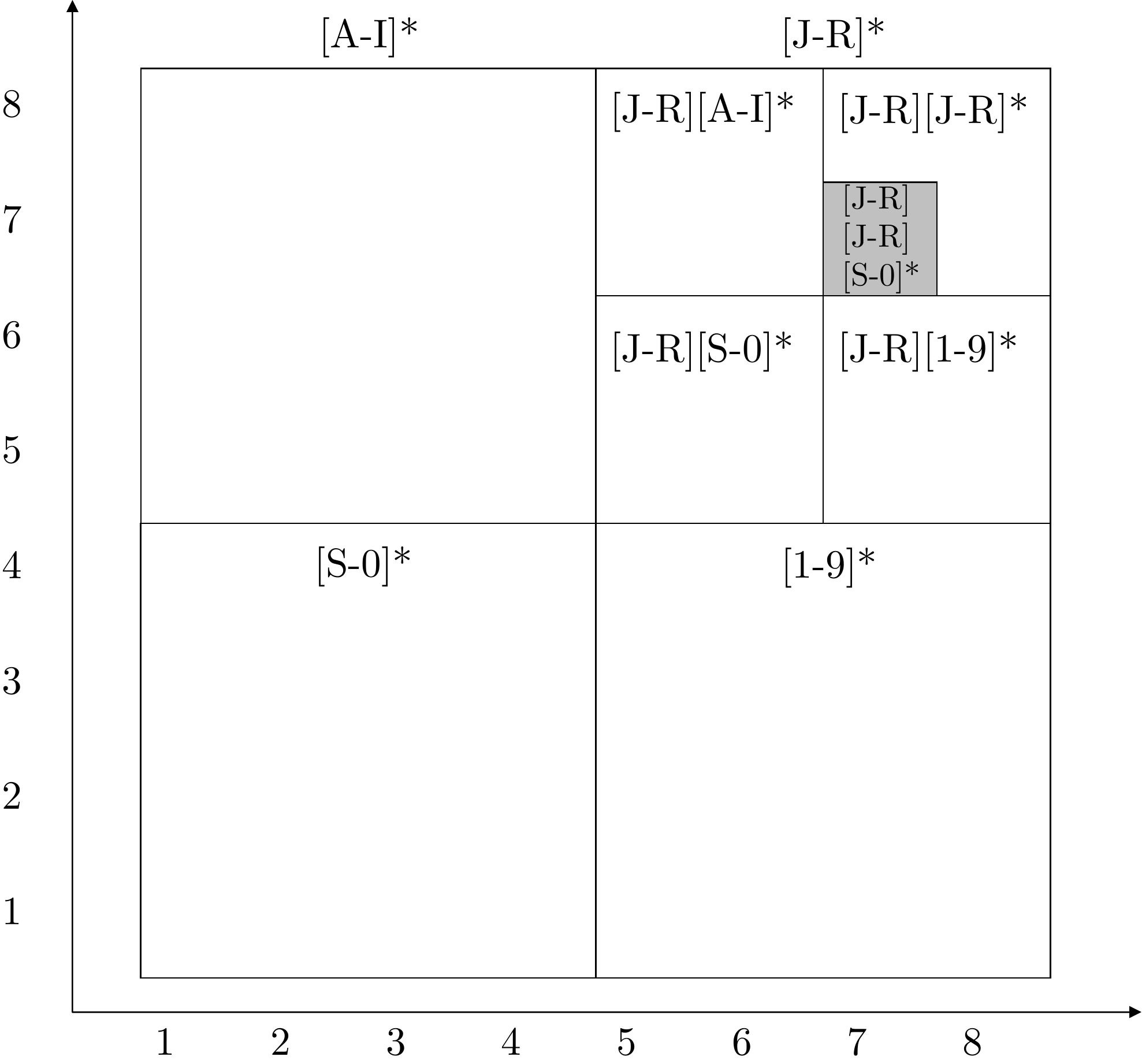}
  \caption{Example quad tree coding}
  \label{fig:quad}
  %\vskip -5mm
\end{figure}

The standard hashing algorithm of the DHT needs to be replaced by a
function which maps equal names to equal places. We use a region quad
tree~\cite{samet1984qar} to preserve the lexical order of all keys.
The result is one is able to spot in advance the area where a set
of keys with the same prefix is located. One example of such a mapping
is shown in Figure~\ref{fig:quad}. The hash space is divided into 4
quadrants, according to the first character of the object name (in the
example, object names can only contain capital letters A--Z and
numbers 0--9; the approach easily generalizes to any character set).
Each of the quadrants is further divided into four quadrants according
to the same mapping, and so on. As a result, any object whose name
starts with the prefix ``JOS'', would be mapped to the shaded area
near the top right corner of the area.

Combined with the DTC algorithm, this mapping enables an efficient
prefix search in a DHT. We simply pick any node at random from the
area of the desired prefix and span a tree over the area covered by
the prefix.

We can also adjust the granularity of the mapping in order to control
how many peers are present in the areas for prefixes of different
lengths. We call this the \emph{split factor} and it works as follows.
As example, consider the mapping shown in Figure~\ref{fig:quad}. The
whole hash space is divided into 4 quadrants and each quadrant is
assigned a set of letters. When objects are stored, we look at the
first character of the object name and select the quadrant which
matches. Then we split that quadrant again into 4 according to the
same rules and map the second character of the object name. Because
the space gets split once per each character, we call this split
factor 1.

Table~\ref{tab:quad-tree-f1} shows how many nodes need to be searched
for a given prefix. (Note that we consider prefixes as characters of
object names, since this would be how users would use a prefix
search.) The numbers have been calculated for a CAN of 1 million nodes
and uniform distribution of peers. As we can see, already relatively
short prefixes of 4--5 characters map to a relatively small number of
peers, on the order of a few thousands. Already with a prefix of 10
characters, we would typically end up with only 1 peer being
responsible.

\begin{table}[!tb]
  \centering
  \begin{tabular}{|l|c|c|c|}
    \hline
    \textbf{Length of}	& \textbf{Nodes} 	 & \textbf{Share of}	& \textbf{Share of}	\\	
    \textbf{Prefix}	& \textbf{in Area} 	 & \textbf{Keyspace}    & \textbf{Keyspace}\\
    		   	&		      	 & \textbf{Split factor 1} &\textbf{Split factor 3}\\
    \hline
    0 & 1,000,000 & 100\%	&100\%\\
    1 & 250,000	  & 25\%	&1.6\%\\
    2 & 62500	  & 6.25\%	&0.02\%\\
    3 & 15625 	  & 1.6\%	&0.0004\%\\
    4 & 3906	  & 0.4\%       &0.0001\%\\ 
    5 &	976	  & 0.1\%	&$<6*10^{-6}\%$\\
    6 &	244	  & 0.02\%&\\
    7 &	61	  & 0.006\%&\\
    8 &	15        & 0.001\%&\\
    9 &	4	  & 0.0004\%&\\
    10&	1	  & 0.0001\%&\\
    \hline
  \end{tabular}
  \caption{Split factor 1 and 1 million nodes}
  \label{tab:quad-tree-f1}
  \vskip -3mm
\end{table}

Increasing the split factor results in a more aggressive splitting.
For example, consider a mapping of letters and digits to 6 bits
(similar to Base64 encoding~\cite{RFC2045}). For each split, we pick 2
bits and map them to one quadrant of the space. One character is 3
such 2-bit groups, and we have to split 3 times before the area of a
single character prefix can be determined. We call this split 
factor 3 and Table~\ref{tab:quad-tree-f1} shows an example.

% \begin{table}[!tb]
%   \centering
%   \begin{tabular}{|l|c|c|c|}
%     \hline
%     \textbf{Length of}	& \textbf{Nodes} & \textbf{Share of}\\
%     \textbf{Prefix}& \textbf{in Area} & \textbf{Keyspace}\\
%     \hline
%     0 & 1,000,000 & 100\%\\
%     1 & 15625	  & 1.6\%\\
%     2 & 244	  & 0.02\%\\
%     3 & 4  	  & 0.0004\%\\
%     4 &	1	  & 0.0001\% \\ 
%     \hline
%   \end{tabular}
%   \caption{Split factor 3 and 1 million nodes}
%   \label{tab:quad-tree}
%   \vskip -5mm
% \end{table}

As we can see in Table~\ref{tab:quad-tree-f1}, increasing the split
factor results in a much more aggressive splitting. For example, a
prefix of 2 characters maps to an area of 244 peers.

We also define split factors smaller than 1, which are less
aggressive than shown in Table~\ref{tab:quad-tree-f1}. This means that
each splitting requires a prefix of multiple characters. For example,
splitting factor 0.5 (i.e., prefix of 2 characters for each split) is
implemented as follows. We arrange all the characters on the x- and
y-axes and then the points in the coordinate space define 2-character
prefixes. We simply divide this space into 4 equal areas and use those
areas for splitting. This approach easily generalizes, thus we can
freely select the granularity of the splitting according to the needs
of the application. We do not change the general overlay structure only the hash function is replaced. 
Thus already known methods for performance optimization~\cite{dabek2004ddl} and 
congestion control~\cite{klemm2006ccd} will apply as well.

\subsection{Range Queries}
\label{sec:range-queries}

Note that regardless of the splitting algorithm and split factor, the
DTC algorithm can span the tree in any convex area of the hash space.
When spanning the tree, the DTC algorithm can also specify
individually how far along each coordinate axis the tree should span.
By selecting the appropriate coordinates, it is possible to span the
tree in such a manner that it corresponds to a range query over the
objects in the DHT. Further evaluation of the practical performance of
range queries is part of our future research.

\subsection{Discussion}
\label{sec:discussion}

A correct split factor may be critical for application performance.
For the case of a prefix search of objects in a DHT, we obviously want
to limit the search areas to as small as possible without unduly
stressing peers. For a high split factor, even short prefixes map to a
very small number of nodes and may thus result in severe load
imbalances because some prefixes are more popular than others. On the
other hand, a high split factor means that only a very small number of
nodes need to be included in the DTC-constructed spanning tree.

Usually the requirements for distributed hash tables demand a more or
less equal distribution of all keys in the hashtable.  On the one
hand, the equal distribution balances the average load per peer which
reduces local hotspots; on the other, the equality destroys the
order of all keys.

One way to reduce hotspots is to choose a smaller split factor for
multidimensional quad trees at the cost of larger search areas. In
addition, there are several approaches for caching or load balancing of
queries within the CAN overlay~\cite{stading:ppc,rao2003lbs}.
Techniques such as virtual nodes\cite{StoicaI:Chord} can also help
alleviate the load imbalance. Evaluating the effectiveness of
different load balancing techniques is part of our future work.

\section{Evaluation}
\label{sec:evaluation}

We now evaluate the performance of our DTC algorithm and compare it
against two other solutions. We tested DTC over two DHTs, Chord and
CAN, to show how its performance in some cases depends on the
underlying DHT. As comparison, we have selected the application-level
multicast on CAN by Ratnasamy et al.~\cite{ratnasamy2001alm} and a
simple flooding scheme, where each node just forwards a message to all
of its neighbors, except the one where it got the message from.

The P2P overlay networks are simulated with the PlanetSim P2P
simulator~\cite{garcia2005pno} and our code will soon be merged into
the main PlanetSim trunk to make it available for others to test.  As
already reported by Jones et al.~\cite{jones2002uce}, implementing CAN
is not always a straight forward process, but we did not encounter any
further problems than mentioned in~\cite{jones2002uce}. We verified
our implementation of the CAN network by comparing it to the results
in~\cite{jones2002uce} and thorough testing on our own, and the
results were as expected. Our Chord implementation is directly based
on the original work in~\cite{StoicaI:Chord}. Our implementation of
the ALM algorithm of~\cite{ratnasamy2001alm} did not face the
race-condition mentioned in~\cite{Castro2003,jones2002uce}, since we
used the cycle based approach of PlanetSim, which prevents the
mentioned race-condition.

We simulated the 4 selected systems in PlanetSim. The two DTC-based
approaches built the spanning tree, as described in
Section~\ref{sec:flooding_algorithm}. ALM from~\cite{ratnasamy2001alm}
and the simple flooding were let run as specified. We varied the size
of the network (values selected according to Section~\ref{sec:hashing}
for reasonable search areas) and, for CAN-based systems, also the
number of dimensions in the CAN. All simulations were repeated 30
times and the reported numbers are averages over the 30 simulation
runs.

We measured the number of messages received by each node and the depth
of the spanning tree. The first metric determines how (in)efficient
the mechanism is and the second determines how quickly search results
are available.

Table~\ref{tab:HopCountsForMulticastInCAN} shows the distribution of
how many messages a given node received in a 2000 node CAN with 10
dimensions. (DTC-Chord was run on a standard Chord of 2000 nodes.) The
table shows for each of the systems how many nodes on average received
the message from the root. We cut the table at 14 messages per node and summed up all the
nodes that received the message 14 times or more (applies only to
simple flooding in this case). The last row shows the total number of
messages sent in the system. Since there are 2000 nodes, 2000 messages
are sufficient in the optimal case.

\begin{table}[!tb]
  \centering
  \begin{tabular}{|c|c|c|c|c|c|c|}
    \hline
    \textbf{Messages} & \textbf{DTC} & \textbf{Simple}   & \textbf{ALM}
    & \textbf{DTC}\\ 
    \textbf{Received} &	\textbf{CAN} & \textbf{Flooding} & &
    \textbf{Chord} \\ 
    \hline			
    0-1 &    2000 & 1 &    1143 &    2000 \\
    2-3 &   					0		&					2				&	684		&0\\
    4-5	&							0		&					6	  		&	78		&0 \\ 
    6-7	&							0		&					22				&	12		&0\\ 
    8-9	&							0		&					67			&	2			&0\\
    10-11&							0		&					247			&	0			&0\\
    12-13&							0		&					925			&	0			&0\\
    $\ge14$&						0		&					7851		&	0			&0\\
    \hline
    Sum& 2000 &	26106 & 3087 & 2000 \\
    \hline						
  \end{tabular}
  \caption{Average number of messages received per node}
  \label{tab:HopCountsForMulticastInCAN}
  \vskip -5mm
\end{table}

As expected, the two DTC-based solutions do not generate any duplicate
messages. ALM performs relatively well, generating about 50\% too many
messages.  We return to the evaluation of the overhead of ALM below.
As Table~\ref{tab:HopCountsForMulticastInCAN} shows, simple flooding
has an extremely high overhead in terms of messages sent. The
factor-13 overhead shown is typical of the performance of simple
flooding.

The performance of the DTC-based algorithms was as expected in all
investigated parameter combinations (e.g., network size, dimensions).
Both of them were able to perform their task always with the
\emph{minimum} number of messages, i.e., as many messages as nodes.

We also evaluated the depth of the spanning tree, since this directly
affects the time it takes to complete the operation (e.g., search or
broadcast). We compared different network sizes from 200 to 20000
nodes and different dimensions in CAN (ranging from 2 to 20).

\begin{figure}[!tb]
  \centering
  \subfigure[5 Dimensional CAN with 20000 nodes]{\label{fig:averagehops-5dim}\includegraphics[angle=-90, width=0.40\textwidth]{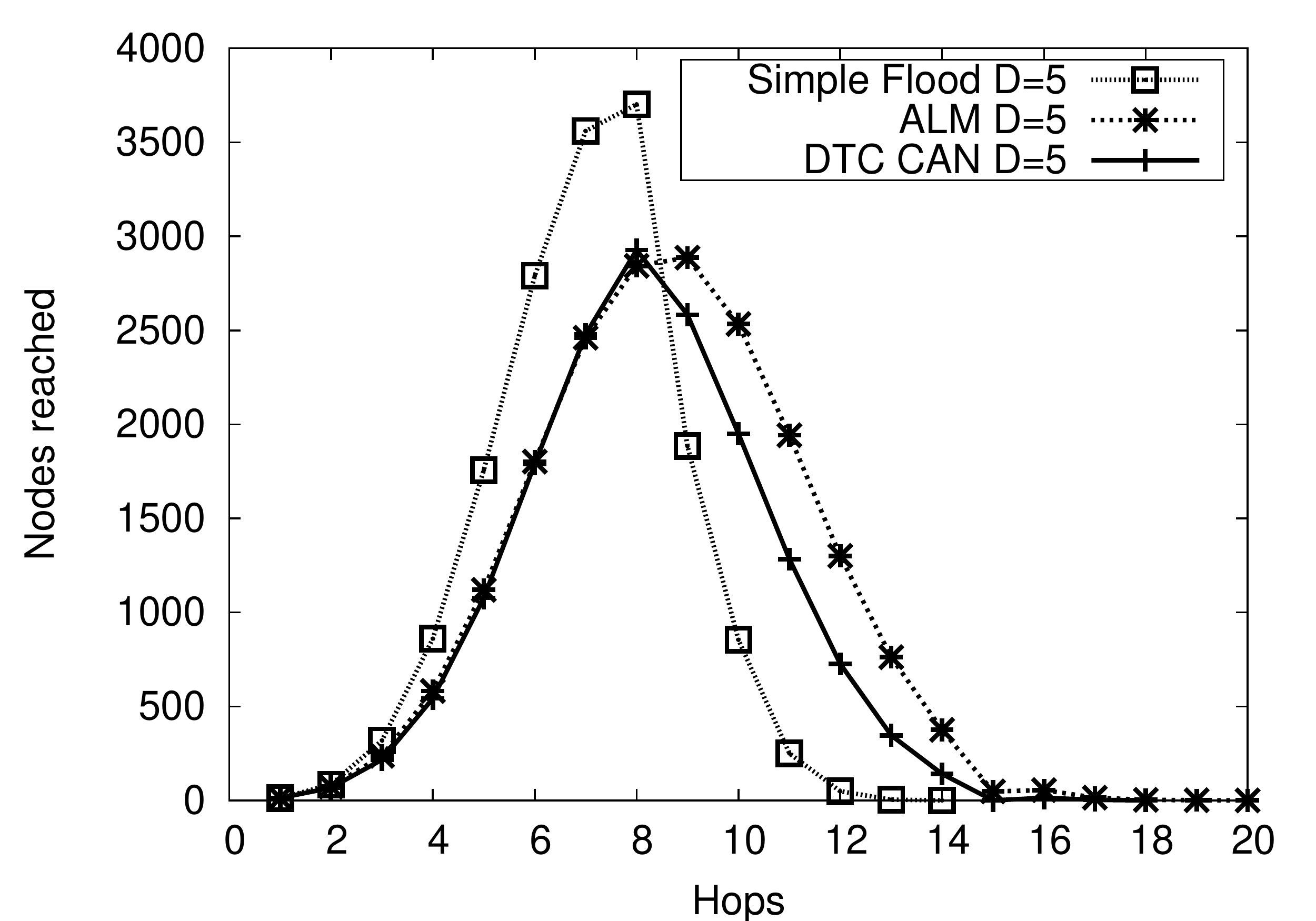}}
  \subfigure[10 Dimensional CAN with 20000 nodes]{\label{fig:averagehops-10dim}\includegraphics[angle=-90, width=0.40\textwidth]{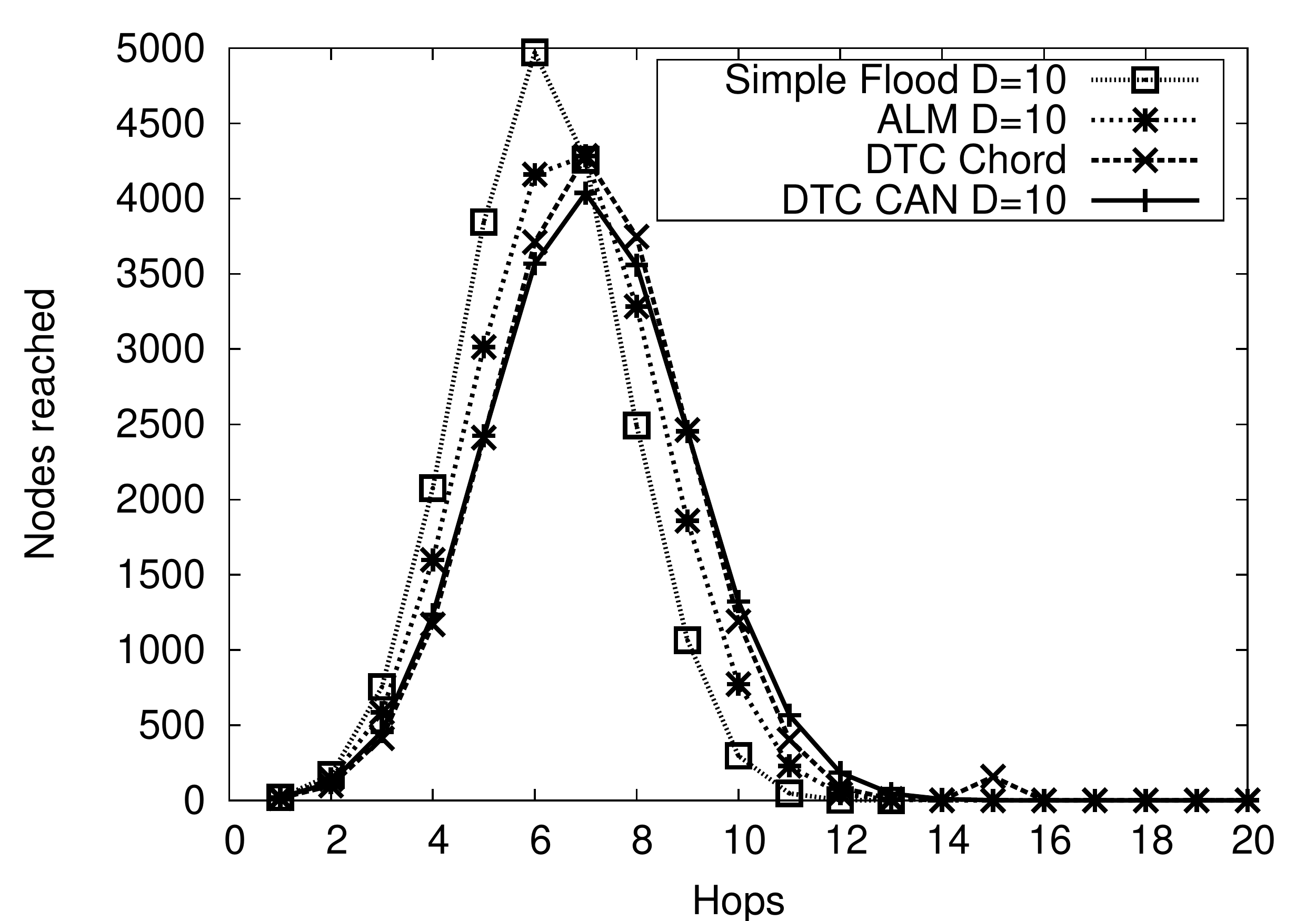}}
  \caption{Query depth}
  \label{fig:query-depth}
  %\vskip -5mm
\end{figure}

Figure~\ref{fig:query-depth} shows 5- and 10-dimensional CANs with
20000 nodes. In Figure~\ref{fig:averagehops-10dim} we also plot
DTC-Chord. The x-axis shows the number of hops and the y-axis shows
how many nodes were reached with that many hops (i.e., how many nodes
are at that depth in the spanning tree).

As we can see, simple flooding has the smallest depth. It always takes
the shortest path to each node, since it forwards a message to all
neighbors but to the sender. Therefore all nodes are reached with the
minimum number of hops.

DTC-CAN and ALM have performance which is slightly lower than simple
flooding and are very close to each other. In the case of the
5-dimensional network and 20000 nodes the optimum would be about 7
hops, while ALM needs 9 and DTC-CAN 8 hops in the average. The greater the number of
dimensions used for the CAN network, the smaller is the difference
between the approaches. Differences in the 10-dimensional CAN network
(see figure \ref{fig:averagehops-10dim}) with 20000 nodes are almost
non-existent; both ALM and DTC-CAN need about 7-8 hops on an average,
while the optimal case would be about 6 hops.

DTC-Chord (shown only in Figure~\ref{fig:averagehops-10dim}) has
performance similar to DTC-CAN and ALM in the 10-dimensional case.
DTC-Chord is independent of the number of dimensions, so it
would be in the same place in Figure~\ref{fig:averagehops-5dim}. The
finger tables of Chord reduce the number of needed hops effectively.
While the CAN network is able to be further optimized by using more
dimensions, the optimum number of hops for Chord is proportional to
the density of the finger tables.

We investigated the performance with dimensions ranging up to 20, but
we did not observe any significant improvement in performance of
DTC-CAN or ALM after 10 dimensions.

\begin{figure*}[!tb]
  \centering
  \mbox{
    \subfigure[Absolute flooding overhead within CAN]{\label{fig:500-20000-messages-dim5-20}\includegraphics[angle=-90, width=0.40\textwidth]{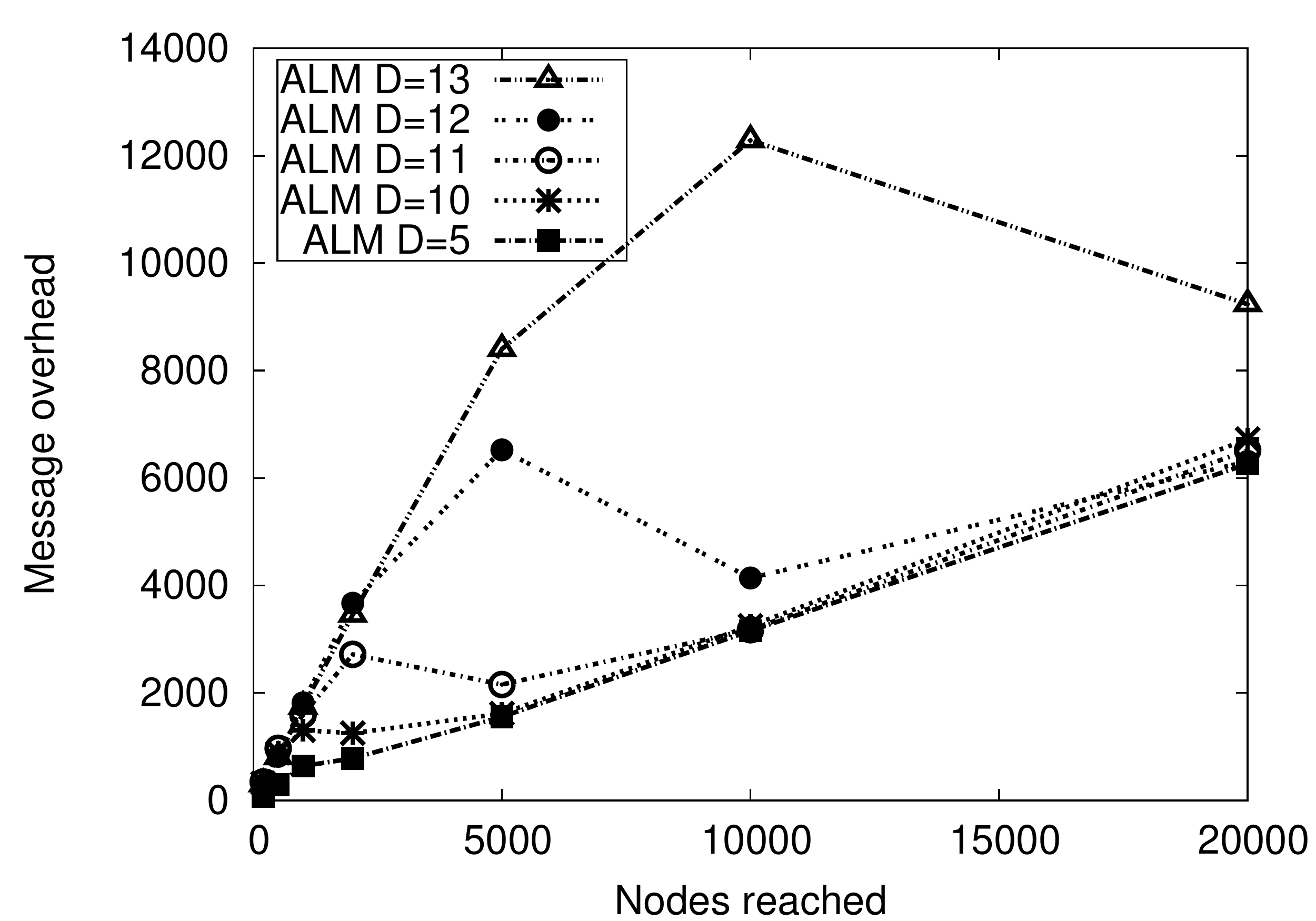}}
    \quad
    \subfigure[Relative flooding overhead within CAN]{\label{fig:500-20000-messages-dim5-15-r}\includegraphics[angle=-90, width=0.40\textwidth]{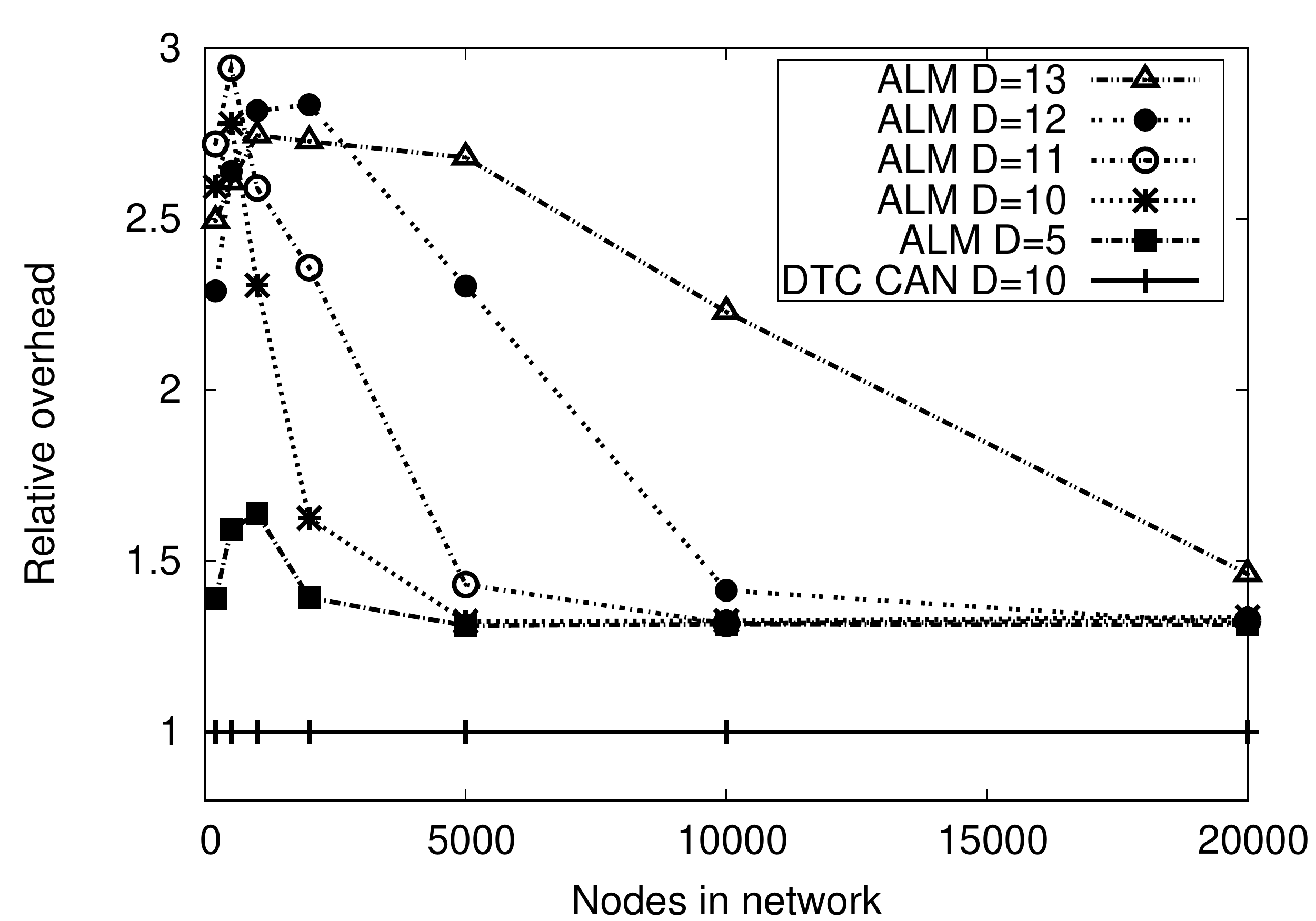}}
    }
  %\caption{Example of DTC-constructed tree}
  %\label{fig:example-dtc-constr}
  %\vskip -5mm
\end{figure*}

We now turn to evaluating the overhead of ALM, which was already shown
in Table~\ref{tab:HopCountsForMulticastInCAN}. We compare ALM against
DTC-CAN. Figures~\ref{fig:500-20000-messages-dim5-20}
and~~\ref{fig:500-20000-messages-dim5-15-r} show how the overhead
evolves as function of network size. The x-axis shows the number of
simulated nodes; we started from 200 nodes and simulated up to 20000
nodes. The y-axis shows the average number of generated messages. We
show several variants of ALM, each with different number of
dimensions. Note that DTC-CAN was always able to perform optimally,
i.e., as many messages as there were nodes.

Figure~\ref{fig:500-20000-messages-dim5-20} shows the absolute
overhead, i.e., how many unnecessary messages ALM sent and
Figure~\ref{fig:500-20000-messages-dim5-15-r} shows the relative
overhead compared to DTC-CAN.

As the network size grows, the relative overhead of ALM in
Figure~\ref{fig:500-20000-messages-dim5-15-r} tends to about
32\%. However, Figure~\ref{fig:500-20000-messages-dim5-20} shows the
interesting behavior of ALM. For every number of dimensions, the curve
has several sharp corners where the overhead changes considerably. The
reason for this is as follows. 

The message overhead is heavily influenced by the absolute number of
dimensions and nodes. The more dimensions, the more nodes are needed
to populate the $d$-dimensional ID space uniformly. As soon as
$\sqrt[d]{n}\ge 2$, the overhead starts to converge to the estimated
32\% (see figure~\ref{fig:500-20000-messages-dim5-15-r}). Below the
critical threshold, the number of duplicate messages steadily
increases with more nodes. This explains the high peaks for the
smaller networks in Figures~\ref{fig:500-20000-messages-dim5-20} and
as Figure~\ref{fig:500-20000-messages-dim5-15-r} shows, the resulting
overhead can be up to 250\%.

In summary, the DTC-based approaches have the advantage of generating
only the minimum amount of traffic, while keeping the depth of the
spanning tree similar to ALM. ALM, on the other hand, has a message
overhead of at least 32\%, in many cases up to 250\%. The simple
flooding approach turns out to be unusable since the enormous message
overhead would cause unacceptable congestions within the overlay.
However, simple flooding has the fastest response time of all compared
approaches.

\section{Robustness of DTC}
\label{sec:robustness-security}

Because the DTC algorithm (and other similar approaches
like~\cite{ratnasamy2001alm,CastroM:SplitStream}) builds a tree, it is
especially vulnerable to defective peers, where a defective peer could
either be the result of a crash or it could be a malicious peer.  If a
peer does not forward the message, then all its children and their
children will not be included in the spanning tree. 

Because all DHTs have mechanisms in place to detect crashed peers and
recover from those crashes, we do not consider such system failures to
be a problem. The only case a message could be lost is if a peer
crashes between receiving a message and forwarding it, which is
extremely rare. In all other cases, we assume that the standard DHT
maintenance takes precedence over the tree spanning messages and that
in such cases, the overlay will first be healed and thus no message
loss or duplication can occur.

In the following, we consider the case of a malicious peer not
forwarding the message correctly onwards. The case of a malicious peer
modifying the message can easily be detected by picking a random
public/private keypair, signing the original message, and including
the public key in the query.
We first evaluated the severity of this problem by selecting a
randomly generated 20000-peer network and spanning a tree with the DTC
algorithm over the whole network. The CAN networks in this test had 10
dimensions.  We varied the fraction of malicious peers and selected
the malicious peers uniformly at random from all the peers. For each
case, we measured the number of peers who were not part of the tree.
Each parameter combination was repeated 30 times and the results we
present are averaged over all the runs.

\begin{figure}[!tb]
  \centering
  \includegraphics[angle=-90, width=0.40\textwidth]{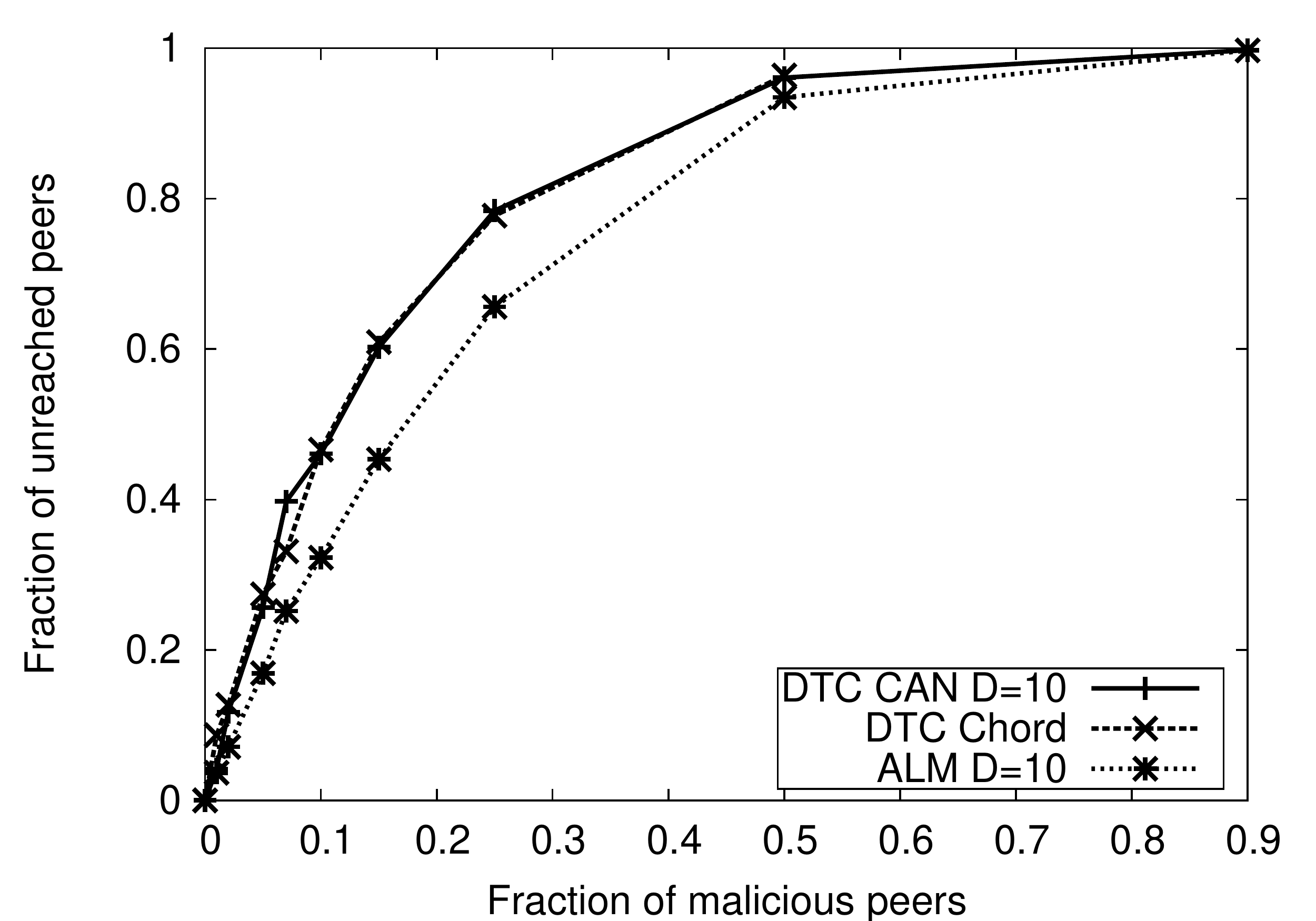}
  \caption{Impact of malicious peers}
  \label{fig:malpeers}
  %\vskip -5mm
\end{figure}

Figure~\ref{fig:malpeers} shows the fraction of unreached peers as a
function of malicious peers for 3 systems from
Section~\ref{sec:evaluation} (DTC-CAN, DTC-Chord, and ALM).  Already
10\% of malicious peers are able to cut off on an average 20\% of the
peers from the spanning tree. Because the DTC algorithm eliminates all
duplicate messages, it is particularly vulnerable to malicious peers,
but the ALM does not fare much better. Only when the fraction of
malicious peers is between 10\% and 50\% does ALM have any marked
improvement over DTC-based systems. Although not shown on
Figure~\ref{fig:malpeers}, the simple flooding approach is extremely
resistant against malicious peers. Peers become unreached only when
the fraction of malicious peers is very high (typically over
70--80\%).

There is an interesting point to note about the effects of the
underlying DHT on the performance of DTC-algorithms. For the case of
1\% malicious peers, Chord-based DTC performed much worse than a
CAN-based DTC. In case of DTC-Chord, 8.6\% of the peers were not
reached, whereas in a DTC-CAN, only 4.2\% of the nodes were unreached.
For comparison, in ALM 3.7\% of the nodes were unreached for that
fraction of malicious peers. For other fractions of malicious peers,
the difference between Chord- and CAN-based DTCs was in practice
negligible, although usually the Chord-based system had the higher
number of unreached peers.

The explanation for this is that in a 10-dimensional CAN, even if a
malicious node is near the root in the spanning tree, it cannot do
much damage. In contrast, if the longest finger of the root in Chord
is malicious, it is able to cut half of the spanning tree. Thus, a
Chord-based DTC is slightly more vulnerable to malicious peers than a
CAN-based DTC.

We now propose solutions for remedying the problems caused by
malicious peers. The fundamental problem is that a single malicious
peer can eliminate the complete sub-tree rooted at that peer from the
spanning tree.  Our primary goal is thus \emph{detecting} the presence
of malicious peers, with as little overhead as possible. The
mechanisms below are not always able to pinpoint the malicious peer;
they only aim at discovering malicious behavior in the tree.

The effectiveness of DTC greatly depends on the kind of
application using DTC. If we are searching for objects stored in
the DHT (as in Section~\ref{sec:hashing}), then the problem
simplifies greatly. This is because all peers should return a response
to the root of the tree indicating whether they have any objects which
match the search. By requiring \emph{all} peers to send a response,
the root can easily check the responses to see whether the complete
area has been searched. Any malicious peer would easily be found out
with such responses.

In the interest of scalability, the responses should be propagated
back along the tree. Furthermore, each peer should sign its response
with a key that is tied to its zone of responsibility in the DHT and
its IP address. These signatures increase the likelihood of
discovering malicious peers, because the malicious peer would have to
invent the zones of responsibility and IP addresses for all the peers
in its subtree. These might overlap with legitimate zones from other
parts of the tree, thus indicating the presence of a malicious peer.
Furthermore, the root could also verify some of the leafs of the tree
(using additional communications) that the responses were actually
sent by them. Although this would reduce the effects of malicious
peers, in the absence of a centralized identity management scheme, it
cannot completely eliminate them.

Another solution is to split the root of the spanning tree and instead
of a single tree, generate several spanning trees, each covering a
small area of the total tree and in such a manner that the union of
the smaller trees covers the whole area. This is easy to perform,
since the DTC algorithm allows us to define the area freely. The
advantage is that each individual search area becomes relatively small
and allows for an easier detection of malicious peers. As discussed in
Section~\ref{sec:hashing}, a prefix length of 4 digits results in a
search area of about 4000 peers. By splitting the query into 10
queries, each query needs to cover about 400 peers. Requiring a
response in such small trees does not present a significant overhead
and thus malicious peers would be easier to detect. Note that the
response traffic is required by the application and is thus
independent of the way the spanning tree is created (DTC, ALM,
flooding, etc.).

The above solutions work well in the case where the tree is spanned
for a purpose where a response is required, e.g., a search. In the
case of a pure one-way tree, such as broadcast, no natural feedback
channel exists. Acknowledgements, similar to the responses in the
search application, would work as described above, but in this case
they would generate additional traffic in the network. However,
without such an acknowledgement mechanism, it is not possible to
detect malicious peers. Thus, we propose to use the same mechanism for
any kind of application. The second solution, the creation of several,
non-overlapping spanning trees, is likely to be more efficient at
detecting malicious peers, but a full evaluation of the proposed
mechanisms is part of our future work.

\section{Related Work}
\label{sec:related-work}

Flooding approaches for P2P networks in general have been extensively
investigated~\cite{jiang2003lef,chawathe2003mgl,TerpstraW:BubbleStorm}.
While simple flooding approaches may apply to unstructured networks,
DHT networks can take advantage of the structured neighbor lists and
reduce the number of duplicate messages.

In the case of CAN, work by Ratnasamy et al.~\cite{ratnasamy2001alm}
implemented an application-level multicast on top of CAN. As our
evaluation shows, this approach can have a significant overhead and
even in the best case, will have an overhead of about 32\%. As we have
shown, the overhead is a function of the network size and CAN
dimensions and in smaller search areas, the overhead can grow
considerably higher. An improved version of ALM is described
in~\cite{Castro2003}.  They reduce duplicate messages, but duplicates
may still occur, especially in the case of uneven zone sizes within
the CAN overlay.

One approach for Range Queries within DHTs~\cite{data-prefix} is to form a tree similar to
our quad tree approach. The difference is that they explicitly use a
DHT in order to store a generated tree for accessing objects. Our
approach, in contrast, does not use the DHT interface for generating a
tree; it directly restructures the mapping from object to nodes. The
advantage is that our solution enables prefix searching without any
overhead.

Another approach for prefix search is presented
in~\cite{10.1109/P2P.2006.24} by Joung et al. The idea is that the 
prefixes build a sub-hypercube within a hypercube which spans a
binomial tree~\cite{10.1109/ICDCS.2005.44}. In general a binomial tree
can be traversed without generating duplicate messages. On the
downside, the hypercube needs to have as many dimensions as bits in
the ID space. Given a common 160~bit ID space, this would lead to 160
dimensions, which in turn leads to 160 neighbors per node for a
binomial tree. Further overhead is generated, because each of the
$2^{160}$ bits needs to be addressable. Therefore a prohibitevely large
number of logical nodes are necessary in smaller networks. In
contrast, our DTC algorithm achieves the same optimal number of
messages with \emph{no overhead} over a standard DHT.

A different approach for prefix searching~\cite{872053}  uses
a combination of different overlay networks, one  for file
management and for site (participants) management.  The
combination of a DHT based overlay (e.g. CAN) with another independent
overlay enables prefix search as well, but at much higher costs. Our
solution does not require any additional maintenance costs or
additional links; all required knowledge can be calculated locally.
While our work concentrates on raw address coding and message
distribution, additional performance optimizations applied to routing
for keyword search in DHT based P2P
networks~\cite{reynolds2003epp,gnawali2002kss} can be used with DTC,
since they are complementary to our solution. We do not alter the
underlying routing scheme and therefore all optimizations used for the
DHT apply to DTC as well.

Applying multi-attribute range queriers to current DHTs has been done by choosing Hubs~\cite{bharambe2004mss}.
The idea of Hubs, which are responsible for one attribute each is independent form the underlying DHT. A hub
can be seen as a separate overlay for each attribute. Our prefix and multicast approach can also be applied to hubs which would lead 
to an optimized ID space for each attribute. On the one hand less nodes would need to be queried, on the other hand a separate overlay network needs to be maintained for each attribute.

\section{Conclusion}
\label{sec:conclusion}

In this paper we have presented a distributed tree construction
algorithm, which is able to create a spanning tree over a part of a
DHT, with no inter-node communication and using only information
available locally on each node. We have also presented a mapping based
on region quad trees for mapping objects to nodes in a DHT in such a
manner that the DTC algorithm is able to perform prefix searches on
content stored in the DHT. Our solution is not limited to searches,
but can also handle broadcast and multicast communications. Our
evaluation shows that DTC performs as expected and the comparison to
similar approaches from literature shows that the message overhead of
existing solutions is considerably higher than the optimal number of
messages sent by DTC. The depth of the spanning trees are similar in
all studied cases. We have also discussed the robustness of DTC
against malicious peers and presented solutions for strengthening DTC.  
Our future investigation will concentrate on evaluating more schemes
for exploiting the structured geometry of DHT overlays for improving
security aspects, multicast performance, and expressiveness of search
queries. We will investigate caching mechanisms to overcome the
potential hotspots which may occur using a location sensitive object
to node mapping.

% \bibliographystyle{IEEEbib}
% \bibliography{semi}

\bibliographystyle{latex8}
%\bibliography{latex8}

\end{document}